\documentclass[twocolumn,showpacs,preprintnumbers,amsmath,amssymb]{revtex4}

\usepackage[dvips]{graphics,color}
\usepackage{epsfig}
\usepackage{times}
\usepackage{graphics}
\usepackage{amsmath}
\usepackage{amsfonts}

\newcommand{\id}{{\sf 1 \hspace{-0.3ex} \rule[0.01ex]{0.12ex}{1.52ex} \rule[.01ex]{0.3ex}{0.12ex} }}
\usepackage{amssymb}
\usepackage{amstext}
\usepackage{float}

\begin{document}



\title{Manipulating quantum information by propagation}

\author{\'Alvaro Perales$^a$}
\email{alvaro.perales@uah.es}

\author{Martin B.\ Plenio$^{b,c}$}%
\email{m.plenio@imperial.ac.uk}

\affiliation{$^a$Dpto. Autom\'atica, Escuela Polit\'ecnica,
Universidad de Alcal\'a, 28871 Alcal\'a de Henares, Madrid,
Spain\\ $^b$Blackett Laboratory, Imperial College London, Prince
Consort Rd, London SW7 2BW, UK\\ $^c$Institute for Mathematical
Sciences, Imperial College London, 53 Exhibition Road, London SW7
2AZ, UK}

\date{\today}

\begin{abstract}
We study creation of bi- and multipartite continuous variable
entanglement in structures of coupled quantum harmonic
oscillators. By adjusting the interaction strengths between
nearest neighbors we show how to maximize the entanglement
production between the arms in a Y-shaped structure where an
initial single mode squeezed state is created in the first
oscillator of the input arm. We also consider the action of the
same structure as an approximate quantum cloner. For a specific
time in the system dynamics the last oscillators in the output
arms can be considered as imperfect copies of the initial state.
By increasing the number of arms in the structure, multipartite
entanglement is obtained, as well as $1 \rightarrow M$ cloning.
Finally, we are considering configurations that implement the
symmetric splitting of an initial entangled state. All
calculations are carried out within the framework of the rotating
wave approximation in quantum optics, and our predictions could be
tested with current available experimental techniques.
\end{abstract}

\maketitle

\section{Introduction}
Generally quantum information processing is considered
theoretically and implemented experimentally as a temporal
sequence of quantum gates applied to quantum bits that are
stationary in space. Many classical information processing devices
function differently however. A microchip usually takes as an
input currents that are propagating through the chip where they
suffer classical logic gates. The emerging current represent the
outcome of the particular classical operation that has been
carried out. In the present work we consider the quantum
information equivalent of this way of processing information, ie
we propagate quantum information through a system of interacting
quantum systems such that, in the course of its propagation
through the system, it suffers a non-trivial quantum state
transformation. In the following we will focus attention on
harmonic chains for their ready availability of exact solutions in
non-trivial settings with many excitations \cite{Eisert P
03,Eisert PBH 03,Plenio HE 04,Plenio S 05,McAneney PK
05,Paternostro KPL 05} but see also recent work in spin-systems in
this direction \cite{Yung B 05,Hadley SB 05}

The aim is to design devices to be integrated in a quantum
information processor in which it will be essential the
availability of reliable sources of entanglement as well as its
transportation and unitary manipulation inside the processor. At
the same time the proposed structures can be considered as cloning
machines of one initial input state.

This paper is structured as follows. In Section \ref{model} we
present the basic physical model and its Hamiltonian as well as
the utilized tools to measure entanglement and its time evolution
in the system. In section \ref{entanglement} we describe the
Y-shaped structure formed by coupled quantum harmonic oscillators,
and fix the initial conditions for entanglement creation. We
present and justify the required tuning of the coupling strengths
between nearest neigbours that creates maximum entanglement
between the last oscillators in the output arms, and provide the
resulting unitary transformation between initial and final states.
We show different results varying the system parameters and
compare them with previous ones and with the upper limit for
entanglement generation in the system. Then we extend the
result to multiarm structures in which multipartite entanglement
is created. Finally we study X-shaped
structures which work as entanglement beam splitters.
In Section \ref{cloning} we study the action of the
same structures as a cloning machines in which the final states in
the last oscillators of the output arms can be considered as
imperfect copies of the initial input state. We provide the
fidelity expression in terms of the squeezing parameter and also
extend the results to multiarm structures which creates multiple
more imperfect copies.

\section{Model and analytical tools} \label{model}
We begin by introducing the basic technical tools that we will be
employing in this work.

\subsection{Hamiltonian}

We begin by briefly describing the model system used in Ref.\
\cite{Plenio HE 04}, a 1D chain of interacting quantum harmonic
oscillators, and the tools used to describe quantum states and
entanglement in it. In the rotating wave approximation (RWA) the
Hamiltonian of the system is given by
\begin{equation}
    \hat{H} =  \sum_{k=1}^{M}\left[
    (1+c)\Bigl({\hat a}^{\dagger}_k {\hat a}_k + \frac{1}{2}\Bigr)
    - c\left(\hat{a}^{\dagger}_{k+1}\hat{a}_k +
    \hat{a}_{k+1}\hat{a}^{\dagger}_k\right)\right].
\end{equation}
We may rewrite this Hamiltonian in terms of position and momentum
operators  ${\hat q}_k$ and ${\hat p}_k$ as is customary in a
large part of the quantum information literature. Defining ${\hat
a} = ({\hat q} +i{\hat p})/\sqrt{2}$ and ${\hat a}^{\dagger}
=({\hat q} -i{\hat p})/\sqrt{2}$ we find
\begin{equation}
    \hat{H} =  \frac{1}{2} \sum_{k=1}^{M}\left[
    {\hat q}_k^2 + {\hat p}_k^2 +
    \frac{c}{2} \left(\hat{q}_{k+1}-\hat{q}_k\right)^2 +
    \frac{c}{2} \left(\hat{p}_{k+1}-\hat{p}_k\right)^2\right],
    \label{HRWA}
\end{equation}
where $c$ is the coupling constant between next neighbors.
In this paper we will use its matrix form
\begin{equation}\label{Hmat}
    \hat{H}=\frac{1}{2} R \left[ \begin{array}{cc} V & 0\\
    0 & T
    \end{array}\right]R^T
    = \frac{1}{2}\sum_{ij=1}^M {\hat q}_i V_{ij} {\hat q}_j
    + {\hat p}_i T_{ij} {\hat p}_j,
\end{equation}
where $R^T = ({\hat q}_1,\ldots,{\hat q}_n,{\hat p}_1,\ldots,{\hat
p}_n)$, and $V$, $T$ are the potential and kinetic matrices
respectively \cite{Audenaert EPW 02}. In the rotating wave
approximation we find $V=T$.

\subsection{Gaussian states}
Analysing the entanglement properties of infinite dimensional
systems is generally technically
involved unless one restricts attention to so-called Gaussian
states \cite{Eisert P 03, Simon SM 87} , which are more easily
described in phase space
introducing the (Wigner-)characteristic function.
Using the Weyl operator $W_{\xi} = e^{i
\xi^{T} R}$ for $\xi\in{\mathbf{R}}^{2n}$
we define the
characteristic function as
\begin{equation}
    \chi_{\rho}(\xi) = \mbox{tr}[\rho W_{\xi}].
    \label{chitr}
\end{equation}
The state and its characteristic function are related to each
other according to a  Fourier-Weyl relation,
\begin{equation}
    \rho = \frac{1}{(2\pi)^n} \int d^{2n}\xi
    \chi_{\rho}(-\xi) W_{\xi} .
    \label{rho}
\end{equation}
Gaussian states are exactly those states for which the
characteristic function $\chi_{\rho}$ is a Gaussian function in
phase space \cite{Simon SM 87} and are completely specified by their
first and second moments, $d$ and $\gamma$
\begin{equation}
    \chi_{\rho}(\xi) =
    \chi_{\rho}(0) e^{-\frac{1}{4}\xi^T \gamma \xi - d^T \xi}.
    \label{chi}
\end{equation}
As the
first moments can be always made zero
utilizing appropriate local displacements in phase space, they are
not relevant in the context of questions related to squeezing and
entanglement and will be ignored in the following. The second
moments can be collected in the real symmetric $2n\times 2n$
covariance matrix $\gamma$ defined as
\begin{eqnarray}
    \gamma_{j,k} &=&
    2 \mbox{Re} \, \mbox{tr}\left[
    \rho \left({\hat R}_j-\langle {\hat R}_j\rangle_{\rho} \right)
    \left({\hat R}_k-\langle {\hat R}_k\rangle_{\rho} \right)
    \right] .
\end{eqnarray}
With this convention, the covariance matrix of the $n$-mode vacuum
is $\gamma=\id_{2n}$.

For a Hamiltonian operator like (\ref{Hmat}) we find that the
covariance matrix of the ground state is \cite{Audenaert
EPW 02}
\begin{equation}
    \gamma = \sqrt{TV^{-1}}\oplus \sqrt{VT^{-1}}
\end{equation}
In RWA we have $T=V$ and the ground state is given by
$\gamma=\id_n\oplus \id_n$, which is the same as the ground-state
of $n$ non-interacting harmonic oscillators.

\subsection{Entanglement of Gaussian States}
As the covariance matrix encodes the complete information about
the entanglement properties of the a Gaussian state, we will use
it in order to quantify the amount of entanglement between two
groups of oscillators. In this work we are going to use two
entanglement measures. When considering pure states, the
appropriate measure of entanglement is the entropy of entanglement
which is defined as the von-Neumann entropy of the reduced density
matrix of one subsystem. In the Gaussian states setting this
reduced density matrix is again described by a covariance matrix
$\gamma_A$  and the displacements. As displacements represent
local unitary basis changes they are irrelevant when considering
the entanglement properties of a Gaussian state. If we denote the
symplectic eigenvalues of $\gamma_A$ describing $m$ harmonic
oscillators by $\mu_i$ we find \cite{Plenio EDC 05} for the
entropy of entanglement the expression
\begin{equation}
    S = \sum_{i=1}^m \left(\frac{\mu_i+1}{2}\log_2\frac{\mu_i+1}{2}  -
        \frac{\mu_i-1}{2}\log_2\frac{\mu_i-1}{2}\right).  \label{Smu1}
\end{equation}
Another measure of entanglement that can also be applied for mixed
quantum states is the logarithmic negativity \cite{Lewenstein HSL
98,Eisert P 99,Lee KPL 00,Eisert Phd,Vidal W 02,Plenio V 05}. This
measure is easy to compute, possesses an interpretation as a cost
function \cite{Audenaert PE 03} and its monotonicity under local
operations and classical communication has been proven recently
\cite{Plenio 05}. Given two parties, $A$ and $B$, that consist of
$m$ and $n$ harmonic oscillators respectively, the logarithmic
negativity is defined as
\begin{equation}
    N(\rho)=\log_2||\rho^{T_B}||_1 = -\sum_{j=1}^{m+n}\log_2(\min(1,|\gamma_j|)),
\end{equation}
where $\rho^{T_B}$ is the partial transposition of $\rho$ with
respect to system $B$, and $||.||_1$ denotes the trace-norm. The
second equality is obtained employing the symplectic eigenvalues
$\gamma_j$ of the covariance matrix $\gamma^{T_B}$ corresponding
to the partially transposed state $\rho^{T_B}$ (see Refs.\
\cite{Audenaert EPW 02,Plenio HE 04}).

\subsection{The equations of motion}
As we are interested in the entanglement evolution we must now
find equations of motion for the covariance matrix. The dynamics
of the covariance matrix under a Hamiltonian quadratic in position
and momentum operators can be obtained straightforwardly from the
Heisenberg equation.
\begin{equation}
    \frac{d}{dt} \hat X(t) = i [{\hat H},\hat X] .
\end{equation}
For our time-independent Hamiltonian Eq.\ (\ref{Hmat}), this leads
to the covariance matrix at time $t$ as
\begin{eqnarray}
\label{num}
\Gamma(t) = \exp^{\left({\left[
\begin{array}{cc}
  0 & T \\
  -V & 0 \\
\end{array}
\right] t}\right)} \Gamma(0)
\exp^{\left({\left[
\begin{array}{cc}
  0 & -V \\
  T & 0 \\
\end{array}
\right] t}\right)}.
\end{eqnarray}
where
\begin{eqnarray}
\Gamma(t) := \left(\begin{array}{cc}
  \gamma_{XX} & \gamma_{XP}  \\
  \gamma_{PX}  & \gamma_{PP}  \\
\end{array}\right)(t).
\end{eqnarray}
These equations of motion will be numerically integrated and, in
some cases, analytically be solved in this paper.

\section{Creating entanglement by propagation} \label{entanglement}
We begin our considerations by devising structures in which
entanglement is generated automatically during the propagation of
quantum information.

\subsection{Bipartite entanglement}
\begin{figure}[h]
\begin{center}
{\resizebox{!}{5cm}{\includegraphics{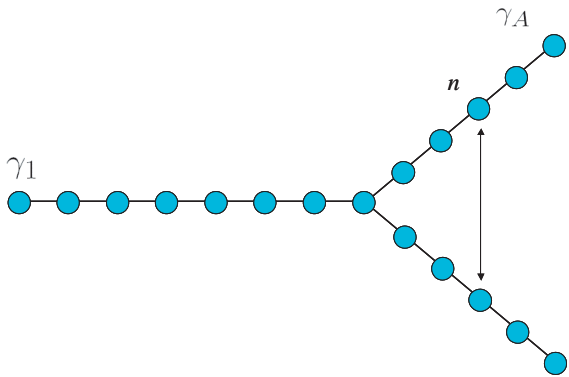}}}
\caption{\label{yshape} Sketch of the Y-shaped structure. A single
input arm consists of $M_{\rm in}$ oscillators (here $8$) is
connected to two output arms each consisting of $M_{\rm out}$
oscillators (here $5$). The arrow indicates how to group
oscillators in the same position $n$ of each branch to apply basis
change (\ref{qQ}).}
\end{center}
\end{figure}

In Ref.\ \cite{Plenio HE 04} a Y-shaped structure as shown in
Fig.\ \ref{yshape} was proposed. One arm, the input arm,
consisting of $M_{\rm in}$ oscillators is connected to two further
arms, the output arms, each consisting of $M_{\rm out}$
oscillators. Only nearest neighbour interactions are considered
and it is assumed that the structure is initially in the ground
state, i.e., at temperature $T=0$. At time $t=0$ we perturb the
first harmonic oscillator in the input arm exciting it either to a
thermal state characterized by covariance matrix elements
$\gamma_{q_1q_1} = \gamma_{p_1p_1} = z$ for some $z$, or a pure
squeezed state characterized by covariance matrix elements
$\gamma_{q_1q_1} = 1/\gamma_{p_1p_1} = z$
\begin{equation}
    \gamma_1(t=0) =\left[\begin{array}{cccccc}
     z & 0 \\
     0 & 1/z   \end{array}
    \right].  \label{gamma1}
\end{equation}

We observe that for an initial thermal state excitation no
entanglement is ever found between the ends of the two arms of the
Y-shape. This can be understood from the observation that a
thermal state is a mixture of coherent states, i.e., displaced
vacuum states. Displacements do not contribute to the entanglement
in a state so that it is sufficient to consider the vacuum state.
If the system is initialized in the vacuum state however it
remains stationary and will evidently not lead to any entanglement
as long as we are considering the RWA. Therefore an initialization
in a thermal state cannot yield entanglement either.

For a squeezed state input on the other hand a considerable amount
of entanglement is generated. These two observations are
resembling closely optical beamsplitters which do not create
entanglement from thermal state input but can generate
entanglement from squeezed inputs (see Ref.\ \cite{Wolf EP 03} for
a comprehensive treatment of the entangling capacity of linear
optical devices).

It is now natural to consider which nearest neighbour coupling
provides the optimal generation of entanglement. These optimal
couplings can be found numerically by adjusting randomly
individual couplings, accepting changes that improve the amount of
entanglement that is being generated in the structure at a given
time for a given input. The results so obtained provide the
intuition by which one then arrives at an analytical argument. In
Ref.\ \cite{Plenio HE 04} it was shown that it is possible to
obtain perfect transmission in a linear chain of $M$ oscillators
if the couplings are adjusted in a square root law
\begin{eqnarray}
    V_{n,n+1} &=& V_{n+1,n} = \frac{c}{2}\sqrt{n(M-n)}  \label{Vsqrt}\\
    V_{n,n}   &=& 1    \label{perfect}.
\end{eqnarray}
With this potential matrix an interchange between the first and the
$M$-th coordinate occurs by waiting for a time $t=\pi/c$ (see Ref.\
\cite{Christandl DEL 03} for an analogous argument in spin
chains).

\begin{figure}[h]
\begin{center}
{\resizebox{!}{6cm}{\includegraphics{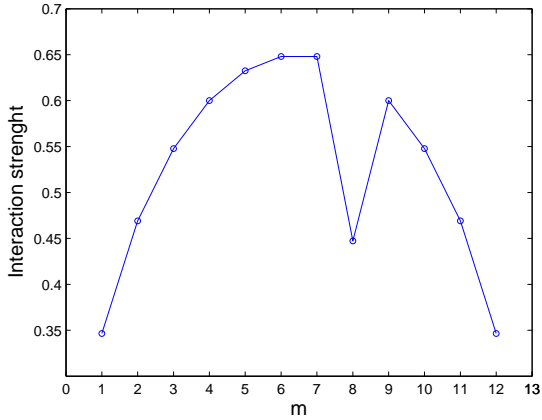}}}
\caption{\label{Vprofile} Square root law profile of the potential
matrix non-diagonal elements which produces maximum entanglement
between both final oscillators in the output arms of the Y-shaped
structure (just one output arm interactions are shown). The
coupling between the last oscillator in the input arm, and the
first oscillator in the output arms must be divided by $\sqrt{2}$.
$M_{\rm in}=8$, $M_{\rm out}=5$ and $c=0.2$.}
\end{center}
\end{figure}

\begin{figure}[h]
\begin{center}
{\resizebox{!}{2.2cm}{\includegraphics{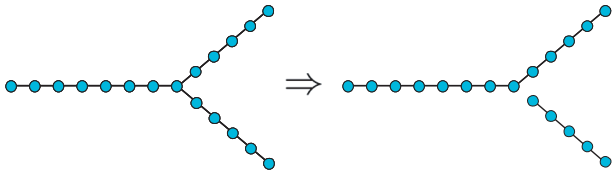}}}
\caption{\label{Trafofigure} Under the transformation (\ref{qQ})
the Y-shape on the left hand side is transformed into a linear
chain consisting of the input arm and one output arm while the
other output arm is now decoupled.}
\end{center}
\end{figure}

\begin{figure}[h]
\begin{center}
{\resizebox{!}{6cm}{\includegraphics{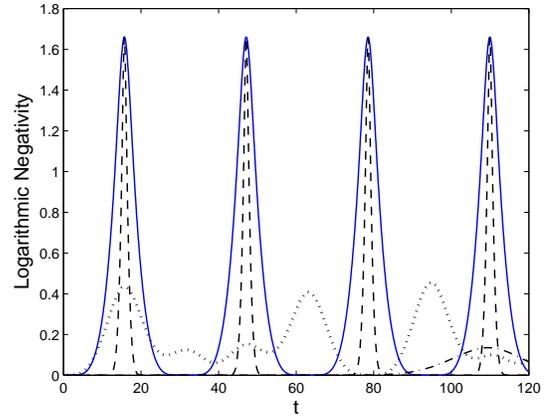}}}
\caption{\label{LogNegT} Time evolution of entanglement between the two
last oscillators in the Y-shaped structure when an initial squeezed state
is prepared in the first oscillator of the input arm. Maximum entanglement
is obtained for\ $t=\pi/c$. Solid line corresponds to $M_{\rm in}=M_{\rm out}=2$
and dashed line to $M_{\rm in}=M_{\rm out}=20$. Dotted and dash-dotted lines
correspond to RWA Hamiltonian with constant couplings for $M_{\rm in}=M_{\rm out}=2$
and $M_{\rm in}=M_{\rm out}=10$, respectively. Coupling constant
is $c=0.2$ and squeezing parameter is $z=10$. }
\end{center}
\end{figure}

To obtain the equivalent perfect transmission distributed between
the arms in the Y-shaped structure, we must modify the coupling
strengths between the oscillator in the junction and the first two
oscillators in the arms, $V_{M_{\rm in},M_{\rm in}+1}$, in the
following way
\begin{equation}
    V_{M_{\rm in},M_{\rm in}+1} = \frac{c}{2}\sqrt{M_{\rm in}(M-M_{\rm in})/2}  \label{Vjunc}.
\end{equation}
In Fig. \ref{Vprofile} these couplings are shown for $M_{\rm
in}=8$ and $M_{\rm out}=5$.
To understand the origin of this choice of coupling strengths we
link it to perfect transmission in a linear chain. Indeed, with a
basis change between both systems it can be proved that the choice
of coupling for the Y-shape corresponds to perfect transmission in
a linear structure consisting of the incoming arm and one outgoing
arm while the other outgoing arm is becoming decoupled (see Fig.
\ref{Trafofigure}). Labelling the two arms with $A,B$ we take
oscillators by pairs in the same position $n$ of each arm, and
make the following change just in arms (see Fig.\ \ref{yshape})
\begin{equation} \label{qQ}
    {\hat q}_n^{A} = \frac{1}{\sqrt{2}}({\hat Q}_n^A + {\hat Q}_n^B)
     , \;\;\;\;\;
    {\hat q}_n^{B} = \frac{1}{\sqrt{2}}({\hat Q}_n^A - {\hat Q}_n^B),
\end{equation}
with analogous equations for momentum operators ${\hat p}_n$.
Introducing this change in the expression for the system
Hamiltonian  (\ref{HRWA}) we obtain that $A$ arm  gets the
perfect transmission coupling strengths Eq.\ (\ref{Vsqrt}),
while $B$ arm is decoupled.
How does this allow us to obtain the optimal couplings for the
y-shape? First, we consider the transformed y-shape depicted on
the right hand side of fig. \ref{Trafofigure}. We perform the same
procedure carried out for the linear chain (see Ref.\ \cite{Plenio
HE 04} for an analytical proof of optimality) to obtain the
optimized couplings for this situation. Transforming back to the
original coordinates, ie the situation on the left hand side of
Fig. \ref{Trafofigure}, we then obtain the coupling strengths
suggested in eqs. (\ref{Vsqrt}-\ref{Vjunc}). This then implies
that the initially squeezed state is, in the transformed picture,
transmitted perfectly to the end of the outgoing arms. Inverting
the transformation, which is indeed the same transformation as
that implemented by a 50/50 beamsplitter, we then obtain a
two-mode squeezed state in the original picture. The results in
\cite{Wolf EP 03} imply that the transformation corresponding to a
50/50 beamsplitter is the optimal linear optics entangling
transformation for an initial squeezed state. This implies that
the procedure as a whole, ie the choice of couplings, is optimal
-- at least in the Gaussian setting. If the initial quantum state
of the first oscillator is squeezed the final states in the last
oscillators in the present maximum entanglement at that time, as
can be observed in Fig.\ \ref{LogNegT}, where logarithmic
negativity is shown. The only effect of increasing the number of
oscillators (adjusting couplings accordingly) is to narrow the
peaks without decreasing the maximum entanglement. This is
contrasted by the behaviour of a system in which all the couplings
between nearest neighbours are equal. Then the magnitude of the
entanglement decreases as the number of oscillator increases, and
irregular time patterns are obtained \cite{Plenio HE 04}.

\begin{figure}[h]
\begin{center}
{\resizebox{!}{7cm}{\includegraphics{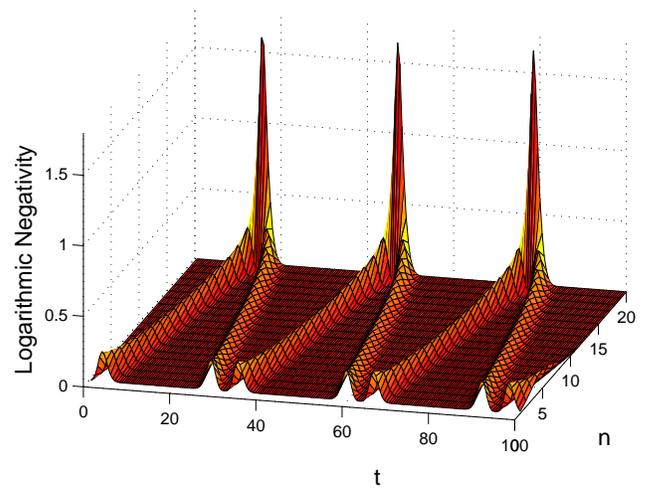}}}
\caption{\label{LogNegTMout} Same as Fig.\ \ref{LogNegT} showing as well
propagation of entanglement along the output arms between oscillator in
the same position of each arm. $M_{\rm in}=2, M_{\rm out}=20$, $c=0.2$ and $z=10$. }
\end{center}
\end{figure}

In Fig.\ \ref{LogNegTMout} we can see how entanglement propagates through the
output arms in the Y-shaped structure. For each pair of oscillators occupying
the same position $n$ in each arm we measure entanglement at each time. Tuning
couplings as in Eqs.\ (\ref{Vsqrt}-\ref{Vjunc}) makes possible maximum
entanglement production between the last oscillators in each arm for $t=\pi/c$.\\

Perfect transmission means that any dispersion has been reversed
in the system at $t=\pi/c$ and all the supplied energy is
accumulated in the two oscillators, one at each end of the
outgoing arms $A$ and $B$, with the remaining oscillators in the
ground state. Therefore this final joint state with $4\times4$
covariance matrix $\gamma_{AB}=\gamma_A \oplus \gamma_B$ is pure,
which is witnessed by the following equality \cite{Eisert P 03}
\begin{equation}
-[\gamma_{AB}(t=\pi/c) \sigma ]^2=\id,
\end{equation}
where $\sigma$ is the $4\times4$ symplectic matrix given by
\begin{equation}
    \sigma =\left[\begin{array}{cccccc}
    0        & {\id }_2\\
    -{\id }_2 &     0    \end{array}
    \right].  \label{sigma}
\end{equation}

At this perfect transmission instants the evolution of the input
state toward the final one $\gamma_1 \rightarrow \gamma_{AB}$ is
an unitary transformation. Its expression can be obtained by first
considering the covariance matrix of the transformed system in
which there is perfect transmission between initial state and A
arm, while B arm is decoupled. In the new coordinates $Q, P$ (we
take out the operator symbols for the sake of symplicity), the
joint state of the last two oscillators in the arms $A,B$
\begin{equation}    \gamma_{AB}^{QP} =
\left[\begin{array}{ccccccc}
    \langle {Q^A}^{2}\rangle & \langle Q^AP^A\rangle   &  0   &  0  \\
    \langle P^AQ^A \rangle & \langle {P^A}^2\rangle  &  0   &  0  \\
    0     & 0    & \langle {Q^B}^2\rangle  & \langle Q^BP^B\rangle \\
    0     & 0    & \langle P^BQ^B \rangle & \langle {P^B}^2\rangle
\end{array}
\right], \label{compare1}
\end{equation}
must therefore equal the direct sum of the squeezed initial state
(perfect transmission with A arm) and the ground state (decoupled
B arm )
\begin{equation}  \gamma_{AB}^{QP} =  \gamma_{1\id}
= \gamma_1 \oplus \gamma_\id =
\left[\begin{array}{ccccccc}
    z & 0   & 0 & 0  \\
    0 & 1/z & 0 & 0  \\
    0 & 0   & 1 & 0  \\
    0 & 0   & 0 & 1
\end{array}
\right]. \label{compare2}
\end{equation}
Comparing the two matrices in Eqs. (\ref{compare1}) and
(\ref{compare2}) we then find
\begin{eqnarray}
\langle {Q^A}^{2}\rangle &=& z, \;\; \langle{P^A}^{2}\rangle = 1/z, \;\;
\langle {Q^B}^{2}\rangle = \langle {P^B}^{2}\rangle = 1  \nonumber \\
\langle Q^AP^A\rangle &=& \langle P^AQ^A\rangle = \langle Q^BP^B\rangle =
\langle P^BQ^B\rangle = 0.
\label{QQPP}
\end{eqnarray}
The covariance matrix of the output oscillators in both arms of
the original coupled Y-shaped structure
\begin{equation}
   \gamma_{AB}^{qp}= \gamma_A \oplus \gamma_B =
    \left[\begin{array}{ccccccc}
    \langle {q^A}^2\rangle & \langle q^Ap^A\rangle   &  0   &  0  \\
    \langle p^Aq^A \rangle & \langle {p^A}^2\rangle  &  0   &  0  \\
    0     & 0    & \langle {q^B}^2\rangle & \langle q^Bp^B\rangle \\
    0     & 0    & \langle p^Bq^B \rangle & \langle {p^B}^2\rangle
    \label{first}
\end{array}
\right],
\end{equation}
can then be obtained by applying (\ref{qQ}). Taking into account
the symmetry ($\gamma_A =\gamma_B$) to notice that $\langle
Q^AQ^B\rangle =\langle P^AP^B\rangle =\langle Q^AP^B +
Q^BP^A\rangle=0$, and with Eqs.\ (\ref{QQPP}) we find
\begin{equation}
   \gamma_{AB}= \gamma_A \oplus \gamma_B = \frac{1}{2}
    \left[\begin{array}{ccccccc}
    z+1 & 0     &  0  & 0  \\
    0   & 1/z+1 &  0  & 0  \\
    0   & 0     & z+1 & 0  \\
    0   & 0     & 0   & 1/z+1
    \label{second}
\end{array}
\right].
\end{equation}
This finally leads to the transformation mapping the initial into
the final state for the Y-shape given by
\begin{equation}
\gamma_A(t=\pi/c) = \gamma_B(t=\pi/c) = \frac{1}{2}[\gamma_1(t=0) + \id]
\label{Relcovs}
\end{equation}

For these pure states we can use entropy of entanglement
Eq.(\ref{Smu1}) as a good measure of entanglement \cite{Eisert P
03,Plenio V 05}. To obtain the dependence of $\mu_1$ with the
squeezing parameter we need to determine the positive eigenvalues
of the matrix \cite{Audenaert EPW 02}
\begin{equation}
    B = -i \sigma \gamma_A,  \label{Bsymp}
\end{equation}
where $\sigma$ is the symplectic matrix (\ref{sigma}). We introduce
Eq.\ (\ref{Relcovs}) (with $\gamma_1$ given by (\ref{gamma1})) in Eq.\ (\ref{Bsymp}),
and take just the positive one of the two real conjugated eigenvalues of $B$
\begin{equation}
   \mu_1 =  \frac{1}{2}\sqrt{z+1/z+2},  \label{mu1}
\end{equation}
which can be introduced in (\ref{Smu1}) with $m=1$ to obtain the expression $S(z)$.\\

It is interesting now to compare the obtained entanglement to the
largest possible entanglement the system can yield given the total
energy $E_T(z)$ that have been supplied through squeezing of the
input oscillator
\begin{equation}
 E_T(z) = \frac{1}{2} ( {\langle q_1^2 \rangle} +
    {\langle p_1^2 \rangle} )-1 = \frac{1}{2}(z + 1/z) -1,
    \label{ET}
\end{equation}

where the $-1$ stands for a shift in energy such that the energy
of coherent states be zero, $E_T(z=1)=0$. Therefore we want to
maximize entropy of entanglement
\begin{equation}
  S(\rho_A) = -\mbox{tr}(\rho_A \log\rho_A) = -\sum_{i=1}^\infty \lambda_i
                                \log \lambda_i ,
 \end{equation}

with the following constraints for probability
and energy conservation
 \begin{eqnarray}
   \mbox{tr}(\rho_A)&=& \sum_{i=1}^\infty \lambda_i = 1 \\
   \mbox{tr}(\rho H)&=& \sum_{i=1}^\infty \lambda_i 2E_i = E_T(z).
\end{eqnarray}

\begin{figure}[h]
\begin{center}
{\resizebox{!}{6cm}{\includegraphics{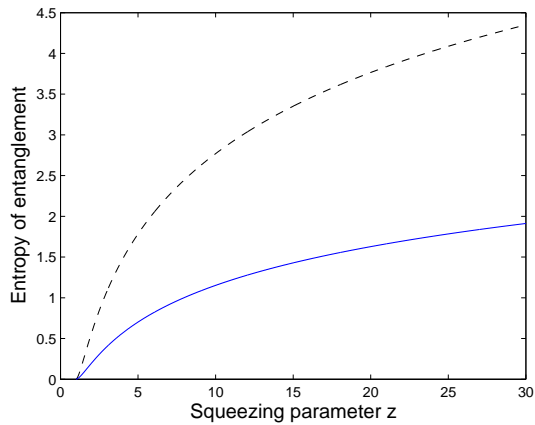}}}
\caption{\label{Sro} (Solid line) Entanglement obtained in
the Y-shaped structure between the oscillators in the extremes of
the output arms (maximum at $t=\pi/c$) and (dashed line) maximum
entanglement in the system given a energy $E(z)$, vs. squeezing
parameter $z$.}
\end{center}
\end{figure}

The $\lambda_i$ are the Schmidt coefficients of the density matrix
$\rho= \sum_{i=1}^\infty \lambda_i |i\rangle \langle i|$ (where we have taken
into account symmetry between particles, i.e.
$|i_A\rangle = |i_B\rangle = |i\rangle$),  and $E_i=i$
are the dimensionless eigenstates of one harmonic oscillator. Using
Lagrange multipliers we obtain the maximum entanglement the system
can yield in terms of the squeezing
parameter $z$
\begin{multline}
S_{max}(\rho_A) = \frac{1}{\mbox{ln}2}
   \left[ \frac{E_T(z)}{2}  \mbox{ln}\left(\frac{2}{E_T(z)}+1 \right) \right. \\
   + \left. \mbox{ln}\left(\frac{E_T(z)}{2}+1 \right) \right] , \label{Smax}
\end{multline}

where the total energy is given by Eq.\ (\ref{ET}). Fig.\ \ref{Sro}
shows the obtained entanglement (maximum in time evolution) in the Y-shaped
structure, Eq. (\ref{Smu1}), and the maximum one, Eq. (\ref{Smax}), in terms
of $z$. Both quantities are independent on the coupling constant $c$ and
on the number of oscillators.

\subsection{Multipartite entanglement}

\begin{figure}[h]
\begin{center}
\vspace{0.6cm}
\includegraphics[viewport= 30 10 500 650,width=5cm, angle=-90]{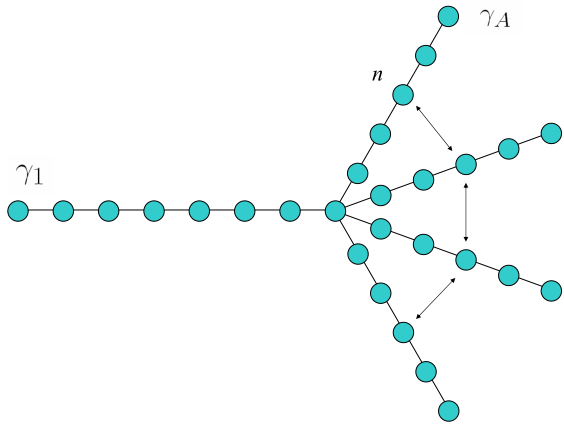}
\vspace{0.3cm}
\caption{\label{multiarm} Multiarm structure for $N_A=4$, $M_{\rm
in}=8$ and $M_{\rm out}=5$. Arrows indicates how to group
oscillators in the same position $n$ of each branch to apply DFT
basis change (\ref{DFT}).}
\end{center}
\end{figure}

The arguments presented above can be extended easily to a
structure in which one has $N_A$ output arms structure as depicted
in Fig.\ \ref{multiarm} (see \cite{McAneney PK 05} for a treatment
of similar structures).

\begin{figure}[h]
\begin{center}
{\resizebox{!}{6cm}{\includegraphics{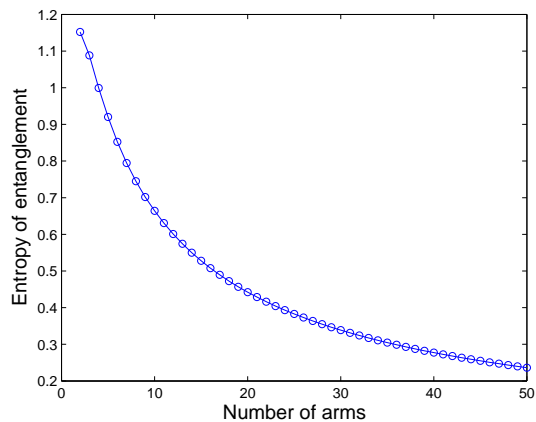}}}
\caption{\label{LogNegN} Multipartite entanglement obtained in
the multiarm structure (maximum in entanglement evolution at $t=\pi/c$) vs.
the number of arms $N_A$.}
\end{center}
\end{figure}

Now the coupling in the junction, Eq. (\ref{Vjunc}), must be
\begin{equation}
    V_{M_{\rm in},M_{\rm in}+1} = \frac{c}{2}\sqrt{n(M-n)/N_A} , \label{VjuncM}
\end{equation}
and the basis change which decouples all arms except one (for which
perfect transmission is obtained) is now a discrete Fourier transform (DFT).
We apply it to position ${\hat q}_n^k$ and momentum
${\hat p}_n^k$ operators of oscillators occupying the same position $n$
in each arm (see Fig.\ \ref{multiarm}) to obtain the new basis operators
\begin{eqnarray}
    {\hat Q}_n^j = \frac{1}{\sqrt{N_A}} \sum_{k=1}^{N_A-1}
    {\hat q}_n^k e^{2\pi i j k/N_A},
    \label{DFT}
\end{eqnarray}

with an analogous expression for momentum operators. We can now follow
the same procedure that yielded Eq.\ (\ref{Relcovs}) to obtain the unitary
transformation which links initial squeezed state in the input oscillator
with the final state at $t=\pi/c$ of the last oscillator in one of the $N_A$
arms
\begin{equation}
\gamma_A(t=\pi/c) = \frac{1}{N_A}[\gamma_1(t=0) + (N_A-1)\id)]  \label{RelcovsNA}.
\end{equation}
which reduces to Eq.\ (\ref{Relcovs}) for $N_A=2$. The initial state is
distributed between $N_A$ final states and we compute multipartite
entanglement as the bipartite entanglement between the oscillator
in the extreme of one arm and the other oscillators in the rest of arms.
Following the same procedure to obtain Eq.\ (\ref{mu1}) we can get now
the positive symplectic eigenvalue of the reduce covariance matrix
in the multiarm structure
\begin{equation}
   \mu_1(N_A) =  \frac{1}{N_A}\sqrt{(z+N_A-1)(1/z+N_A-1)},  \label{muNA}
\end{equation}
which it is introduced in Eq.\ (\ref{Smu1}) to obtain the expected
behaviour shown in Fig. \ref{LogNegN}:
multipartite entanglement decreases with increasing number of arms.

\subsection{Beamsplitters}
Extending arguments of previous section we now study X-shaped
arrays. Such structures may be used to split the entanglement in a
state. In qubit systems, entanglement splitting cannot be achieved
without loss \cite{Buzek VPKH 97} but we will see that the
Gaussian setting permits lossless splitting.

To obtain the equivalent perfect transmission, coupling strengths
must be adjusted analogously to the Y-shaped case, Eqs.\
(\ref{Vsqrt}, \ref{perfect}), but with an interaction applied to
the coupling between the central oscillator and its four nearest
neighbours given by
\begin{equation}
    V_{M_{\rm in-1},M_{\rm in}} = V_{M_{\rm in},M_{\rm in}+1} =
     \frac{c}{2}\sqrt{M_{\rm in}(M-M_{\rm in})}. \label{Vjuncx}
\end{equation}

\begin{figure}[h]
\begin{center}
{\resizebox{!}{6cm}{\includegraphics{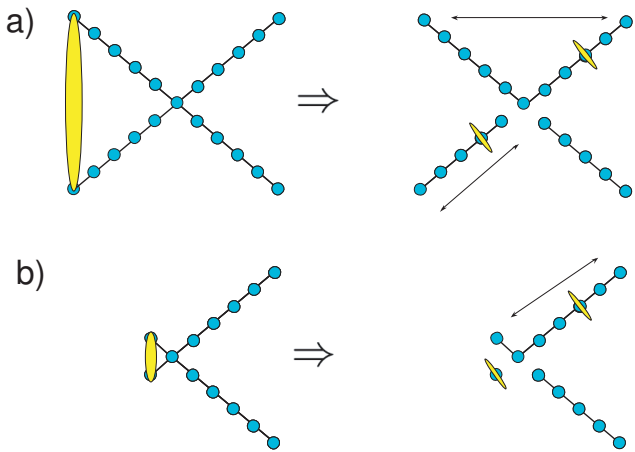}}}
\caption{\label{Trafox} {\bf a)} Under the transformation Eq.
(\ref{qQ}) the X-shape on the left hand side is transformed into a
linear chain and two decoupled arms. The initial two mode squeezed
state (represented by the yellow disk across two arms) is then
transformed into a tensor product of two single mode squeezed
states (individual yellow disks) . One of these can move through
all the structure, while the other one remains in the input arm,
giving rise to the complex time-space pattern shown in Fig.\
\ref{X20}. {\bf b)} When there is just one oscillator in the input
arm, one of the states always remains in the first oscillator and
entanglement is always present in the input arm as shown in Fig.\
\ref{X1}}.
\end{center}
\end{figure}

\begin{figure}[h]
\begin{center}
{\resizebox{!}{7cm}{\includegraphics{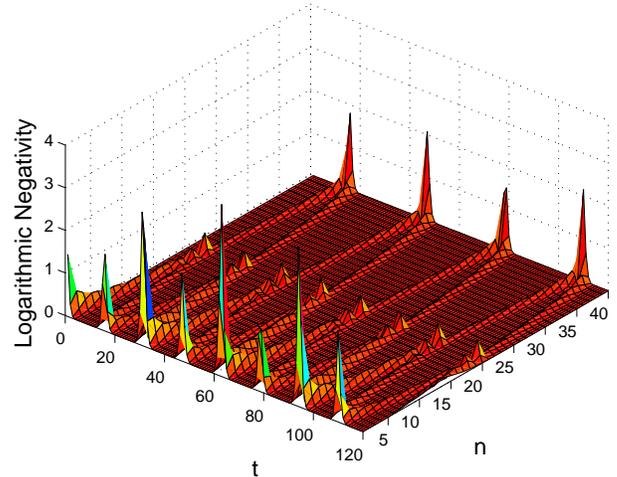}}}
\caption{\label{X20} Time-space evolution of entanglement in the X-shaped structure.
For times when one of the squeezed states is in the output arm, regular transmission
patterns are obtained because the other one must stay in the input arm.
(see Fig.\ \ref{Trafox}a). However when the state returns to the input arm irregular
patterns are obtained as now both states are in the input arms. Maximum entanglement
is obtained when both squeezed states are in the input oscillator.}
\end{center}
\end{figure}

\begin{figure}[h]
\begin{center}
{\resizebox{!}{7cm}{\includegraphics{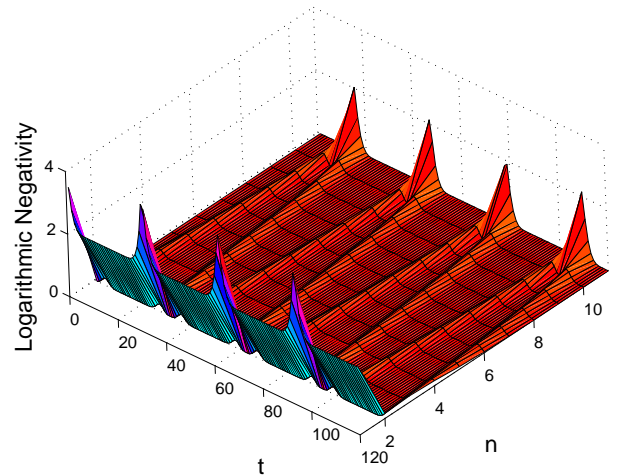}}}
\caption{\label{X1}Time-space evolution of entanglement in the X-shaped structure
when there is just one oscillatr in the input arm (see Fig.\ \ref{Trafox}b).
There is always entanglement between the two oscillators in the input arm
as one of the squeezed states remains always in it. }
\end{center}
\end{figure}

As initial condition we now create an entangled two-mode squeezed state
between the first two oscillators in the input arm, labelled with $1,2$
\begin{eqnarray}
    \gamma_{q_1 q_1} &=& \gamma_{q_2 q_2} = \gamma_{p_1 p_1}
    = \gamma_{p_2 p_2} = \cosh(r) \nonumber \\
    \gamma_{q_1 q_2} &=& - \gamma_{p_1 p_2} = \sinh(r),
\end{eqnarray}
with $\cosh(r)=z$. The resulting time-space evolution of
entanglement is shown in Figs.\ \ref{X20}, \ref{X1}, and we sketch
an explanation for the observed patterns in the caption of Fig.\
\ref{Trafox}.

\section{Cloning} \label{cloning}
At the times when perfect transmission is obtained adjusting the
couplings, the states at the end of both arms can be considered as
approximate copies of the input states. However, we must be
careful to adjust the diagonal element of the $V$ and $T$ matrices
to optimize quality of the clones as the local rotation of the
individual oscillators must now be taken into account. The reason
is that we are now comparing actual quantum states at different
times, $\rho_1(t=0)$ and $\rho_A(t=\pi/c)$, while in entanglement
calculation the local rotations did not affect the amount of
entanglement. We find that in Eq.\ (\ref{perfect}) we must modify
diagonal elements in the following way
\begin{equation}
V_{n,n}= \left\{ \begin{array}{cccccc}
   c/2 \;\;\;  &\mbox{for}& \;\;\; M_{\rm in}+M_{\rm out} \;\;\; \mbox{even}\\
     c \;\;\;  &\mbox{for}& \;\;\; M_{\rm in}+M_{\rm out} \;\;\; \mbox{odd} \end{array}
    \right.,
\end{equation}
which can be expressed as
\begin{equation}
    V_{n,n}= \frac{1}{4}[3-(-1)^{(M_{\rm in}+M_{\rm out})}]c.
\end{equation}
It is worthwhile pointing out that with these and any other
arbitrary value for the diagonal elements we also obtain optimum
entanglement creation, as internal degrees of freedom are equal
for botharms and henceforth entanglement is not affected. 


The fidelity between the {\em pure} input state $\rho_1$ and one
of the two copies, described by the reduced density operator
$\rho_A$, is given by
\begin{equation}
    F = \langle \Psi_1 | \rho_A | \Psi_1 \rangle = \mbox{tr}(\rho_1 \rho_A).
    \label{Fid}
\end{equation}

We now insert in this expression the input state
$\rho_1 = |\Psi_1\rangle\langle\Psi_1|$
given by Eq.\ (\ref{rho}), and consider no displacement ($d=0$) and initial
state $\chi_\rho(0)$ equal to one in Eq.\ (\ref{chi}). Using expression
(\ref{chitr}) and solving the resulting Gaussian integral we obtain
\begin{equation}
    F(\gamma_1, \gamma_A) = \frac{2}{\sqrt{\mbox{det}(\gamma_1 + \gamma_A)}},
    \label{Fid0}
\end{equation}
where $\gamma_1$ as the initial
covariance matrix of the input oscillator (\ref{gamma1}), and
$\gamma_A$ is the final one at $t=\pi/c$ of the A-arm output
oscillator ($\gamma_A = \gamma_B$ for symmetry).

\begin{figure}[h]
\begin{center}
{\resizebox{!}{6cm}{\includegraphics{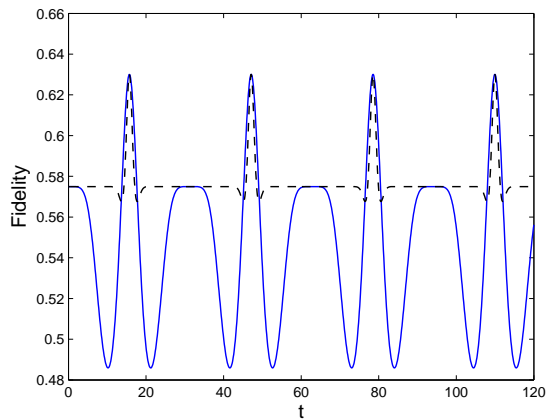}}}
\caption{\label{FidT} Time evolution of fidelity between initial squeezed
state in the
first oscillator in the input arm and final state in the last oscillator
of the output arm in the Y-shaped structure. Maximum fidelity is
obtained for\ $t=\pi/c$. Solid line corresponds to $M_{\rm in}=M_{\rm out}=2$
and dashed line to $M_{\rm in}=M_{\rm out}=20$. $c=0.2$ and $z=10$. }
\end{center}
\end{figure}

\begin{figure}[h]
\begin{center}
{\resizebox{!}{6cm}{\includegraphics{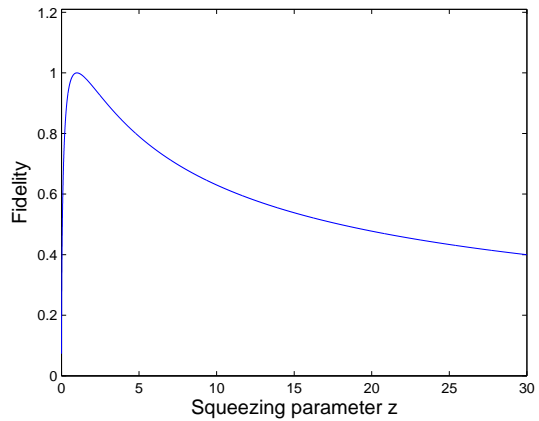}}}
\caption{\label{FidZ} Fidelity obtained in
the Y-shaped structure between the first oscillator in the input arm and the
last oscillator in the extreme of one of the output
arms (maximum at $t=\pi/c$), vs. squeezing parameter $z$.}
\end{center}
\end{figure}

In Fig.\ \ref{FidT}
the fidelity time evolution is shown. The value $F\approx0.575$ corresponds
to overlap between input squeezed state $\gamma_{q_1q_1} = 1/\gamma_{p_1p_1} = z$
and ground state $\gamma_A = \id $. At the
same time for which we obtained perfect transmission and maximum
entanglement, $t=\pi/c$, we obtain as well maximum fidelity. At this
times rest of oscillators are at ground states and relation (\ref{Relcovs})
between covariance matrices of input and output oscillators holds. If we insert
this relation in Eq.\ (\ref{Fid0}) and express $\gamma_1$ elements
in terms of the squeezing parameter $z$, Eq.\ \ref{gamma1}, we get
\begin{equation}
    F(z) = \frac{4}{\sqrt{3z+3/z+10}},
    \label{Fidz}
\end{equation}
which is depicted in Fig. \ref{FidZ}. For $z=1$ we have a ground state
as input and a ground state as output and so the maximum is reached at $F(1)=1$
and fidelity decreases as squeezing is increased.

\begin{figure}[h]
\begin{center}
{\resizebox{!}{6cm}{\includegraphics{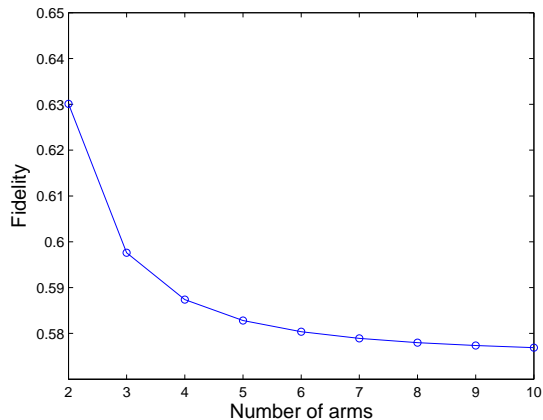}}}
\caption{\label{FidN} Fidelity in a 1 to $N_A$ cloner obtained in
the multiarm structure (maximum in fidelity evolution at $t=\pi/c$) vs.
the number of arms $N_A$, for $z=10$.}
\end{center}
\end{figure}

In the $N_A$ multiarm structure the fidelity can be calculated inserting the unitary
transformation (\ref{RelcovsNA}) in the fidelity expression (\ref{Fid0}) to get
\begin{equation}
    F(z,N_A) = \frac{2N_A}{\sqrt{(N_A^2-1)(z+1/z)+ 2N_A^2+2}},
    \label{FidzNA}
\end{equation}
which reduces to Eq.\ (\ref{Fidz}) for $N_A=2$. In Fig.\ \ref{FidN}
we can see that with increasing number of arms we obtain a faster
decreasing of fidelity than the one obtained for entanglement. For
$N_A\rightarrow\infty$ fidelity approaches the value $F\approx0.575$
which corresponds to an initial state with $z=10$ and a final ground state.

\section{Conclusions}
We have considered situations in which we propagate quantum
information through a system of interacting quantum systems such
that, in the course of this propagation, it suffers a non-trivial
quantum state transformation. In particular we have considered
entanglement creation, quantum cloning and entanglement splitting
in systems of quantum harmonic oscillators both analytically and
numerically under the rotating wave approximation (RWA) as it is
appropriate in a quantum optical setting.

We have presented geometrical structures that allow to create, by
propagation, the complex unitary operations required for the above
processes. Such hard wired pre-fabricated structures may also be
'programmable' by external actions to add further functionality to
them. Therefore quantum information would be manipulated through
its propagation in these devices somewhat analogous to modern
micro-chips and as opposed the most presently suggested
implementations of quantum information processing where stationary
quantum bits are manipulated by a sequence of external
interventions such as laser pulses.

All these investigations were deliberately left at a device
independent level. It should nevertheless be noted that there are
many possible realizations of the above phenomena. These include
nano-mechanical oscillators \cite{Eisert PBH 03}, arrays of
coupled atom-cavity systems, photonic crystals, and many other
realizations of weakly coupled harmonic systems, potentially even
vibrational modes of molecules in molecular
quantum computing \cite{Tesch R 02}.
We hope that these ideas may lead to the development of novel ways
for the implementation of quantum information processing in which
the quantum information is manipulated by flowing through
pre-fabricated circuits that can be manipulated from outside.

{\em Acknowledgements:} Support by a Royal Society Incoming
Visitors award is gratefully acknowledged. This work is part of
the QIP-IRC (www.qipirc.org) supported by EPSRC (GR/S82176/0) and
has also been supported by the EU Thematic Network QUPRODIS
(IST-2001-38877), the Leverhulme Trust grant F/07 058/U and
Project UAH-PI12005/68 of the University of Alcal\'a.

\end{document}